\documentclass[%
reprint,
nobibnotes,
amsmath,amssymb,
prb,
floatfix,
]{revtex4-2}

\usepackage{graphicx}
\usepackage{dcolumn}
\usepackage{bm}
\usepackage{xcolor}
\usepackage{amsmath}
\usepackage{mathtools}
\usepackage{braket}
\usepackage[caption=false, position=top]{subfig}
\usepackage{lipsum}
\usepackage{hyperref}
\usepackage{booktabs}
\usepackage{array}
\usepackage{multirow}
\usepackage{longtable}
\usepackage{siunitx}
\DeclareSIUnit{\angstrom}{\text{\AA}}
\usepackage{physics}

\makeatletter
\def\nobreakhline{%
  \noalign{\ifnum0=`}\fi
    \penalty\@M
    \futurelet\@let@token\LT@@nobreakhline}
\def\LT@@nobreakhline{%
  \ifx\@let@token\hline
    \global\let\@gtempa\@gobble
    \gdef\LT@sep{\penalty\@M\vskip\doublerulesep}
  \else
    \global\let\@gtempa\@empty
    \gdef\LT@sep{\penalty\@M\vskip-\arrayrulewidth}
  \fi
  \ifnum0=`{\fi}%
  \multispan\LT@cols
     \unskip\leaders\hrule\@height\arrayrulewidth\hfill\cr
  \noalign{\LT@sep}%
  \multispan\LT@cols
     \unskip\leaders\hrule\@height\arrayrulewidth\hfill\cr
  \noalign{\penalty\@M}%
  \@gtempa}
\makeatother

\newcolumntype{P}[1]{>{\centering\arraybackslash}p{#1}}
\newcolumntype{M}[1]{>{\centering\arraybackslash}m{#1}}

\usepackage{float}
\usepackage{afterpage}  

\begin{document}
\newcommand{\GG}{\textbf{G}}
\newcommand{\qqq}{\textbf{q}}
\newcommand{\rr}{\textbf{r}}
\newcommand{\kk}{\textbf{k}}
\newcommand{\ee}{\text{e}}

\preprint{APS/123-QED}

\title{BSE+ calculations for 2D materials: a unified description of excitons and plasmons}

\author{Amalie H. Svaneborg$^{1}$}
\email{amhso@dtu.dk}
\author{Kristian S. Thygesen$^{1}$}

\affiliation{$^1$CAMD, Department of
Physics, Technical University of Denmark, DK - 2800 Kongens Lyngby,
Denmark}

\date{\today}

\begin{abstract}
The Bethe-Salpeter equation (BSE) accurately describes low-energy optical spectra in materials
with strong excitonic effects, but its high computational cost limits the number of electron-hole transitions that can be included. Neglecting high-energy transitions leads to an underestimation of the real part of the dielectric function, often producing a spurious plasmon peak in the BSE electron energy loss spectrum (EELS). The recently introduced BSE+ method addresses this by combining a four-point BSE-like equation for the irreducible polarisability with a two-point Dyson equation that includes the high-energy transitions at the Random Phase Approximation (RPA) level. Here, we present a detailed account of the method, extend it to two-dimensional materials, and apply it to a set of transition metal dichalcogenide monolayers. BSE+ preserves the excitonic features of the BSE at low energies while reproducing the plasmonic structure of the RPA at higher energies, yielding good agreement with experimental EELS data across the full energy range and strongly suppressing the spurious plasmon. BSE+ converges much faster than BSE with respect to the electron-hole basis size, at a comparable computational cost, and is implemented in the GPAW code.
\end{abstract}

\keywords{}
\maketitle

\section{Introduction}
Accurate theoretical predictions of the optical spectra and dielectric properties of solids require proper treatment of the two-particle electron-hole (e-h) interaction. Within the independent-particle picture, the optical response is described by the Random Phase Approximation (RPA), which captures the essential features of the dielectric function of bulk metals and moderately screened semiconductors. However, the RPA fails to reproduce the bound-state peaks in the absorption spectrum for materials that exhibit pronounced excitonic features, such as systems with weak dielectric screening including insulators and two-dimensional (2D) materials \cite{spataru2004excitonic, huser2013dielectric}.

The Bethe-Salpeter Equation (BSE) accounts for the attractive direct screened interaction between an excited electron and the hole left behind \cite{onida2002electronic,onida1995ab,albrecht1998ab,benedict1998theory,rohlfing1998electron,marini2009yambo,yan2012optical}. It can yield accurate optical spectra, even when excitonic features are pronounced. Within the static approximation of the direct screened interaction, the BSE can be rewritten as an eigenvalue problem for a two-particle Hamiltonian. The number of matrix elements scales as $(N_vN_cN_k)^2$, where $N_{v/c}$ is the number of valence/conduction bands used to construct the e-h basis, and $N_k$ is the number of $k$-points. Constructing and diagonalizing the BSE Hamiltonian quickly becomes computationally intractable. In practice, we cannot include enough bands and $k$-points to fully converge the dielectric function. Several methods exist to reduce the practical cost, including approaches that avoid the explicit summation over empty states \cite{rocca2012solution}, efficient diagonalization schemes \cite{fuchs2008efficient} and iterative Lanczos-based solvers \cite{gruning2011}, along with double-grid and Wannier interpolation methods \cite{kammerlander2012speeding, gillet2016efficient, alliati2022double}. 

Yet, the fundamental limitation remains: only low-energy excitations can be included in the BSE e-h basis without exhausting the computational resources. This is sufficient to describe the excitonic features and the imaginary part of the dielectric function near the absorption onset. However, the Kramers-Kronig relations imply that the real part cannot be converged without including transitions far above the band edge.

Alternatively, the two-point formalism of time-dependent density functional theory (TDDFT) can be used \cite{runge1984density}, where an exchange-correlation kernel $f_{xc}$ encodes the excitonic effects. The two-point structure of $f_{xc}$ renders the Dyson equation computationally tractable compared to the four-point BSE. The simplest choice, $f_{xc}=0$ recovers the RPA. Going beyond the RPA, the long-range correction kernel of the form $f_{xc}=-\alpha/q^2$ \cite{reining2002excitonic} is computationally efficient but requires determining a system-dependent parameter $\alpha$. The parameter-free bootstrap kernel \cite{sharma2011bootstrap} constructs $f_{xc}$ self-consistently from the dielectric function itself. However, while it yields reasonable excitonic spectra for some semiconductors, it performs poorly for wide-gap insulators with strongly bound excitons \cite{buyn20217}. The nanoquanta kernel \cite{sottile2003parameter} is derived directly from the BSE, and can reproduce BSE-quality absorption spectra, but at a computational cost comparable to the BSE itself. Despite these advances, no single two-point kernel has proven universally reliable across a diverse range of materials, while remaining computationally affordable. 

We have previously introduced BSE+ \cite{bseplus}, which improves the convergence of the BSE by including transitions outside the e-h subspace at the RPA level. We constructed an irreducible polarisability within the limited e-h transition space of the BSE and embedded it into an RPA-like Dyson equation that is unconstrained in the number of bands. BSE+ has a computational cost comparable to the BSE but converges much faster with respect to the size of the e-h basis. As a result, it yields a dielectric function that accurately accounts for the low-energy excitonic effects while achieving RPA-level band convergence at all energies. Applying it to a set of bulk semiconductors and insulators, we demonstrated that BSE+ accurately captures the low-energy excitonic peaks, while yielding a much better description of the static refractive index compared to both BSE and RPA. 

In the present paper, we give a more detailed account of the BSE+ method. We derive the underlying equations more thoroughly, make the approximations explicit, and discuss the implications and limitations of the method. We further demonstrate that BSE+ can be applied to 2D materials. These systems exhibit strong excitonic effects that make the RPA inadequate, and often require dense $k$-point sampling that makes the BSE difficult to converge. Having a response function that simultaneously accounts for excitonic effects and converges across the full energy range, paves the way for future total energy calculations via the adiabatic connection fluctuation dissipation theorem, where both low- and high-energy contributions to the correlation energy are important \cite{olsen2014static}. BSE+ is implemented in the GPAW code, and supports both bulk and 2D calculations.

\section{Theoretical framework}
In this section, we first establish the general formalism of the dielectric function, then treat the special case of 2D materials, and finally derive the BSE+ equation, which improves the convergence of the BSE density response function $\chi(\omega)$.

The density response function describes the first-order change in electronic density $\delta n$ induced by a perturbation in the external potential $\delta v_{\text{ext}}$. A wide range of macroscopic observables can be derived from it, or equivalently from the dielectric function $\epsilon(\omega)$. Expressed in a basis of plane waves, the two are related through

\begin{equation}
    \varepsilon_{\GG\GG'}^{-1}(\qqq,\omega)=\delta_{\GG\GG'}+V_{\GG}(\qqq)\chi_{\GG\GG'}(\qqq,\omega),
\label{eq:eps_GG}
\end{equation}

where $\GG$ are the reciprocal lattice vectors, $V_{\GG}$ is the Coulomb interaction, and $\qqq$ is a vector in the first Brillouin zone. The macroscopic dielectric function is then obtained from the $\GG = \GG' = 0$ component of the inverse dielectric matrix, 

\begin{equation} 
\varepsilon_{\mathrm{M}}( \qqq, \omega) = \frac{1}{\varepsilon_{00}^{-1}(\qqq, \omega)}.
\label{eq:eps_M}
\end{equation} 

Within the BSE, the excitation energies
and associated two-particle wave functions needed to construct the density response function is obtained by solving the eigenvalue problem for the two-particle Hamiltonian,

\begin{equation}
    \sum_{S'} \mathcal{H}(\qqq)_{SS'}A^{\lambda}_{S'}(\qqq) = E^{\lambda}(\qqq)A^{\lambda}_{S}(\qqq),
\label{eq:HAEA}
\end{equation}

where $S = \{n,m,\kk\}$ labels the e-h pair index corresponding to a transition from band $n$ at momentum $\kk$ to band $m$ at momentum $\mathbf{k}+\mathbf{q}$. We include spin-orbit coupling throughout, so the single-particle states are two-component spinors and the band indices run over spinor bands. 
The BSE Hamiltonian takes the form, 

\begin{equation}
\begin{split}
    \mathcal{H}_{SS'}(\qqq) = (\varepsilon_{m\kk+\qqq} &-  \varepsilon_{n\kk})\delta_{SS'} \\ &- \left(f_{m\kk+\qqq} - f_{n\kk}\right)K_{SS'}(\qqq),
\end{split}
\label{eq:BSE_Hamiltonian}
\end{equation}

where $f_{n \kk}$ is the occupation of band $n$ with momentum $\kk$, and the e-h interaction kernel is

\begin{equation}
    K_{SS'}(\qqq) = V_{SS'}(\qqq) - W_{SS'}(\qqq),
\label{eq:kernel}
\end{equation}

where $V_{SS'}$ is the bare Coulomb interaction and $W_{SS'}$ is the statically screened e-h interaction. 

The density response function can be expressed in the e-h basis from the solution to Eq.(\ref{eq:HAEA}),

\begin{equation}
\begin{split}
     \tilde{\chi}_{SS'}(\qqq,\omega)= &\sum_{\lambda} A_{S}^{\lambda}(\qqq)A_{S'}^{\lambda}(\qqq)^{*} \\& \times   \left(\frac{f_{m\kk+\qqq} - f_{n\kk}}{\omega-E_{\lambda} +i\eta(\omega) } - \frac{f_{m'\kk'+\qqq} - f_{n'\kk'}}{\omega+E_{\lambda} + i\eta(\omega)} \right) ,
\end{split}
\label{eq:BSE_chi_matrix_eq}
\end{equation}

where $\eta(\omega)$ is a frequency-dependent broadening parameter. Within the Tamm-Dancoff approximation, the eigenvectors $A^{\lambda}(\qqq)$ are orthogonal and the $\lambda$-sum runs only over positive excitation energies.

In a plane-wave basis, the density response function takes the form

\begin{equation}
\begin{split}
    \tilde{\chi}_{\GG\GG'}(\qqq,\omega)= \frac{1}{\Omega} \sum_{SS'}  \rho_{S}(\GG) \tilde{\chi}_{SS'}(\qqq,\omega) \rho^*_{S'}(\GG'),
\end{split}
\label{eq:polarizability_GG}
\end{equation} 

where $\Omega$ is the crystal volume and the associated pair densities are

\begin{equation}
     \rho_{S}(\GG)\equiv \rho_{n\kk}^{m\kk+\qqq}(\GG)=\bra{\psi_{n\kk}}\ee^{-i(\qqq+\GG)\rr}\ket{\psi_{m\kk+\qqq}}.
\label{eq:matrix_elements}
\end{equation}

In the optical limit $\qqq \to 0$, the pair densities reduce to
\begin{equation}\label{eq:opticallimit}
\lim_{\mathbf q \to \mathbf 0}\rho_{n\kk}^{m\kk+\qqq}(\mathbf 0)=\frac{i\mathbf q\cdot \langle \psi_{n\kk}|\nabla|\psi_{m\kk}\rangle}{\varepsilon_{n\kk}-\varepsilon_{m\kk}}.
\end{equation}

Since $V_{0} \propto 1/q^2$, the $q$-dependence cancels in the dielectric function in Eq. (\ref{eq:eps_GG}). 

Several experimentally accessible spectroscopic quantities follow directly from the macroscopic dielectric function. One of them is the electron energy loss spectrum (EELS), which can be determined from

\begin{equation}
\begin{split}
    \text{EELS}(\omega) &= -\frac{4\pi}{q^2}\text{Im}\left(\chi_{\GG=\GG'=0}(\qqq,\omega)\right).
\end{split}
\label{eq:eels_def}
\end{equation}

In bulk materials, Eq. (\ref{eq:eels_def}) is equivalent to $\text{Im}\left(1/\varepsilon_{\mathrm M}(\omega)\right)$. 

\subsection{2D materials}
Since our calculations are based on plane-wave DFT, which assumes periodicity in all three spatial dimensions, 2D materials must be modeled in supercells with a vacuum region separating the periodically repeated layers \cite{2dsystems}. The out-of-plane supercell length is denoted $L$. To suppress interactions between periodic images, the Coulomb interaction must be truncated, and for a 2D geometry, the truncated interaction satisfies $V_{\text{trunc}} \propto 1/q$. The irreducible density response function is constructed from single-particle states as usual, but when building the full interacting response via the Dyson equation, the bare Coulomb interaction must be replaced by the truncated one. 

For a bulk material, the out-of-plane length is fixed by the physical lattice constant and the macroscopic dielectric function is well-defined. For a 2D material in a supercell, however, the large vacuum region artificially enlarges $L$. Since $\Omega = A \cdot L$, where $A$ is the in-plane unit cell area, and the density response function in the e-h basis does not depend on $L$ (adding vacuum does not alter the material's response to an external field) it follows from Eqs. (\ref{eq:eps_GG}) and (\ref{eq:polarizability_GG}) that $\varepsilon_M(\omega)=1+\mathcal{O}(1/L)$. In the limit of large $L$, the spatial average of the microscopic potential is dominated by the vacuum, and $\varepsilon_{\mathrm{M}} \to 1$. The macroscopic dielectric function therefore loses its physical meaning for isolated 2D materials.

The appropriate quantity is instead the 2D density response function $\chi^{\text{2D}}_{\GG\GG'}$ defined to be independent of $L$. In what follows, we define this object and show how it can be computed in practice. For a 2D material it is natural to Fourier transform in the in-plane directions only, decomposing all reciprocal lattice and wave vectors into in-plane and out-of-plane components, $\GG = (\GG_{||}, G_z)$ and $\qqq = (\qqq_{||}, q_z)$. The response function then depends explicitly on the out-of-plane coordinates $z$ and $z'$, yielding $\chi_{\GG_{||}\GG'_{||}}(z,z', \qqq_{||}, \omega)$. Since the electron density is confined to the slab, this quantity vanishes when $z$ or $z'$ lies in the vacuum region, and the 2D density response function is therefore defined by integrating out the out-of-plane dependence \cite{latini_phd},

\begin{equation}
    \chi_{\GG_{||}\GG'_{||}}^{\text{2D}}(\qqq_{||}, \omega) = \int dz\, dz' \chi_{\GG_{||}\GG'_{||}}(z,z', \qqq_{||}, \omega).
\label{eq:chi2d}
\end{equation}

Imposing periodic boundary conditions in $z$ with period $L$ and expanding in discrete plane waves $\ee^{iG_zz}$ with $G_z = 2\pi m/L$, one obtains,

\begin{equation}
    \chi_{\GG\GG'}^{\text{2D}}(\qqq_{||}, \omega) = L \chi_{\GG_{||}\GG'_{||}}.
\end{equation}

Since $\chi_{\GG\GG'}$ in 3D is normalized to the supercell volume $\Omega = A\cdot L$ (Eq. (\ref{eq:polarizability_GG})), it scales as $1/L$ and the explicit factor of $L$ cancels this dependence. Thus, $\chi_{\GG\GG'}^{\text{2D}}$ is computed by constructing the density response function with the truncated Coulomb interaction and multiplying by $L$. In practice, convergence to the isolated-layer result, requires $L$ to be large enough that interactions between periodic images are negligible.

The EELS spectrum is obtained by substituting $\chi^{\text{2D}}_{\GG\GG'}$ directly into Eq. (\ref{eq:eels_def}). Note that the Coulomb truncation is applied only within the Dyson equation, where it removes the interaction between periodic images. The bare Coulomb interaction is retained in Eq. (\ref{eq:eels_def}), which describes the coupling of the external probe to the induced density. Since $\chi^{\text{2D}}_{\GG\GG'}$ is independent of $L$ by construction, the resulting spectrum is free from the supercell artifacts that affect the macroscopic dielectric function. 

\subsection{BSE+}
The effective two-particle Hamiltonian in Eq. (\ref{eq:BSE_Hamiltonian}) contains $N_vN_cN_k$ e-h pair states, so the number of matrix elements scales as $(N_vN_cN_k)^2$. Constructing the kernel requires the statically screened Coulomb interaction $W_{\GG\GG'}(\qqq)$, which is evaluated once for each $q$-point, with each evaluation demanding a full RPA calculation of the dielectric function. After constructing the Hamiltonian, it must be diagonalized, which constitutes the dominant computational cost.

As a consequence, BSE calculations suffer from severe scaling limitations. In practice, this restricts the included bands to states near the Fermi level, so that only low-energy excitations are reliably described, whereas higher-energy transitions are systematically neglected. We denote the set of transitions that can be included due to computational limitations $\mathcal T$.

\begin{figure}[h]
    \centering
\includegraphics[width=\linewidth]{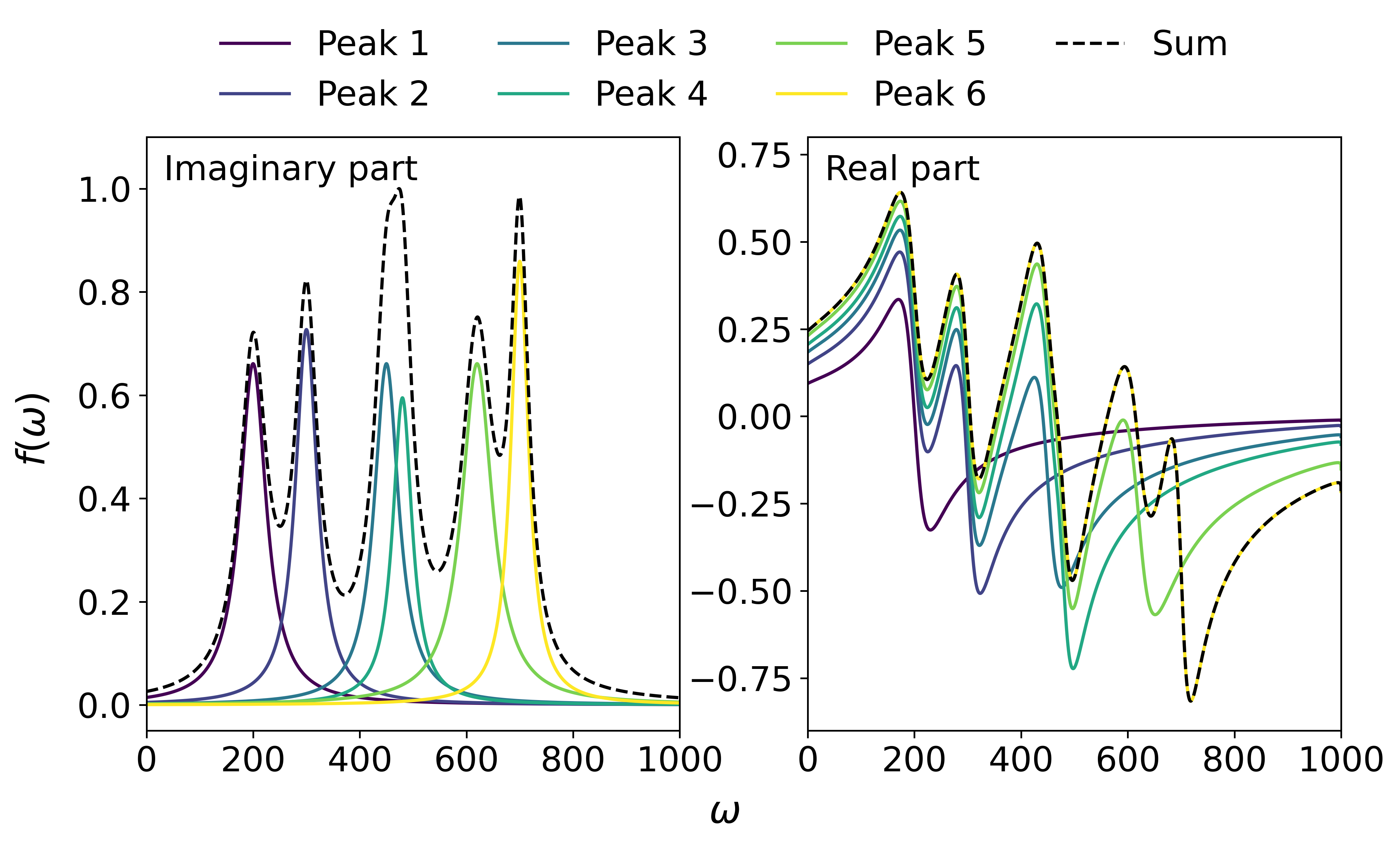}
    \caption{Lorentzian peaks are added successively to the imaginary part (left), with the corresponding real part (right) obtained via  a Hilbert transform. Dashed black lines show the cumulative total.}
    \label{fig:kk}
\end{figure}

A more fundamental limitation arises from the Kramers-Kronig relations, which couple the real and imaginary parts of the dielectric function: The real part depends on the full frequency range of the imaginary part. This is illustrated in Fig. \ref{fig:kk}, where Lorentzian contributions are added sequentially to the imaginary part, and the corresponding real part of the cumulative spectrum is obtained via a Hilbert transformation. Although each Lorentzian is spectrally localized with only a small broadening, their cumulative addition modifies the real part of the dielectric response over the entire frequency range, including the low-energy region. Consequently, incomplete convergence of the high-energy spectrum in BSE calculations leads to an underestimation of the real part of the dielectric function, even at low energies. 

Generally, when performing RPA calculations, we can include as many bands (and $k$-points) as needed to converge the calculation.  However, the RPA lacks excitonic effects. As a result, it fails to accurately reproduce low-energy spectral features dominated by excitons. Although the inclusion of higher-energy transitions improves the convergence of the real part of the dielectric function relative to BSE, the absence of excitonic contributions still leads to an underestimation of the real part of the dielectric response at low energies.

BSE+ combines the strengths of both approaches by capturing the excitonic features of the spectrum as described by the BSE, while achieving a convergence of the density response function comparable to that of the RPA. This improvement is obtained by including transitions outside the active BSE e-h subspace $\mathcal T$ at the RPA level in the irreducible density response function. 

The derivation of the BSE+ equation is based on the ordinary BSE Dyson equation,

\begin{equation}
\begin{split}
    \Tilde{\chi}_{SS'}^{\text{BSE}}(\qqq,\omega)  &= \Tilde{\chi}_{SS'}^{0}(\qqq,\omega)  \\ &+ \sum_{S_1S_2}\Tilde{\chi}_{SS_1}^{0}(\qqq,\omega)K_{S_1S_2}(\qqq)\Tilde{\chi}_{S_2S'}^{\text{BSE}}(\qqq,\omega),
\end{split}
\label{eq:chi_BSE}
\end{equation}

where tildes indicate restriction to $\mathcal T$, $K_{SS'}$ was defined in Eq. (\ref{eq:kernel}) and $\chi^0$ is the non-interacting density response function.

An irreducible density response function is defined by replacing the full kernel with only the screened direct e-h interaction,

\begin{equation}
\begin{split}
    \Tilde{\chi}_{SS'}^{\text{irr}}(\qqq,\omega)  &= \Tilde{\chi}_{SS'}^{0}(\qqq,\omega)  \\ &- \frac{1}{2}\sum_{S_1S_2}\Tilde{\chi}_{SS_1}^{0}(\qqq,\omega)W_{S_1S_2}(\qqq)\Tilde{\chi}_{S_2S'}^{\text{irr}}(\qqq,\omega).
\end{split}
\label{eq:chi_irr}
\end{equation}

Expressed in a plane-wave basis, the BSE Dyson Eq. (\ref{eq:chi_BSE}) becomes

\begin{equation}
\begin{split}
\Tilde{\chi}^{\text{BSE}}_{\GG\GG'}&(\qqq,\omega) =\Tilde{\chi}_{\GG\GG'}^{\text{irr}}(\qqq,\omega) \\& + \sum_{\GG_1\GG_2} \Tilde{\chi}_{\GG \GG_1}^{\text{irr}}(\qqq,\omega)V_{\GG_1\GG_2}(\qqq)\Tilde{\chi}_{\GG_2\GG'}^{\text{BSE}}(\qqq,\omega).
\end{split}
\label{eq:BSE_GG}
\end{equation}

The irreducible response $\tilde \chi^{\text{irr}}_{SS'}$ is extended beyond $\mathcal{T}$ by adding missing contributions at the RPA level, replacing the restricted response $\tilde \chi^0$ by $\chi^0$ in Eq. (\ref{eq:chi_irr}):
 
\begin{equation}
\begin{split}
    \chi_{\GG\GG'}^{\text{irr}}(\qqq,\omega) &= \Tilde{\chi}_{\GG\GG'}^{\text{irr}}(\qqq,\omega) - \Tilde{\chi}_{\GG\GG'}^{0}(\qqq,\omega) + \chi_{\GG\GG'}^{0}(\qqq,\omega),
\end{split}
\label{eq:chi_irr_new}
\end{equation}

where $\chi^{0}_{\GG\GG'}$ can be calculated from
\begin{equation}
\begin{split}
    \chi_{\GG\GG'}^{0}(\qqq,\omega) = \frac{2}{\Omega}&\sum_{\kk, n,m} (f_{n\kk} - f_{m\kk+\qqq}) \\& \times \frac{\rho_{n\kk}^{m\kk+\qqq}(\GG)\rho_{n\kk}^{m\kk+\qqq}(\GG')^*}{\omega+\varepsilon_{n\kk}- \varepsilon_{m\kk+\qqq}+i\eta(\omega)}.
\end{split}
\label{eq:chi0}
\end{equation}

Replacing $\Tilde \chi^{\text{irr}}_{\GG\GG'}$ in Eq. (\ref{eq:BSE_GG}) by $\chi^{\text{irr}}_{\GG\GG'}$ defined in Eq. (\ref{eq:chi_irr_new}) yields the BSE+ equation,

\begin{equation}
\begin{split}
\chi^{\text{BSE+}}_{\GG\GG'}&(\qqq,\omega) =\chi_{\GG\GG'}^{\text{irr}}(\qqq,\omega) \\& + \sum_{\GG_1\GG_2} \chi_{\GG \GG_1}^{\text{irr}}(\qqq,\omega)V_{\GG_1\GG_2}(\qqq)\chi_{\GG_2\GG'}^{\text{BSE+}}(\qqq,\omega).
\end{split}
\label{eq:BSE_plus_polarizability}
\end{equation}

Eq. (\ref{eq:chi_irr_new}) is the main idea in BSE+. Within $\mathcal{T}$, the screened interaction $W$ couples e-h pairs and produces the bound excitonic states characteristic of the BSE. Outside $\mathcal{T}$, the irreducible response reduces to the bare RPA bubble $\chi^0$. Thus, in BSE+, we neglect the action of $W$ among transitions outside $\mathcal{T}$ together with the cross-coupling between transitions inside and outside $\mathcal{T}$. Pairs in $\mathcal{T}$ still couple to pairs outside, but only through the bare Coulomb interaction. Denoting the transitions outside $\mathcal{T}$ by $\mathcal{U}$, the BSE+ kernel can be illustrated as

\begin{equation}
K = 
\vcenter{\hbox{$
\begin{array}{r@{\,}c}
 & \smash{\raisebox{+12pt}{$\hspace{8pt}\begin{matrix}
   \overbrace{\rule{0pt}{0pt}\hspace{2.7em}}^{\textstyle\mathcal{T}} & \hspace{-0.4em}\overbrace{\rule{0pt}{0pt}\hspace{0.9em}}^{\textstyle\mathcal{U}}
   \end{matrix}$}} \\[-10pt]
\begin{matrix} \mathcal{T}\, \lbrace \\[0pt] \mathcal{U}\, \lbrace \end{matrix} &
\begin{pmatrix} V - W & V \\[3pt] V & V \end{pmatrix}
\end{array}
$}}
\end{equation}

Neglecting $W$ outside $\mathcal{T}$ is justified by the energy scales involved. Near the band edge, the transition energies are small and comparable to the exciton binding energy, allowing $W$ to produce bound states. Far above the gap, the transition energy is much larger than the binding energy that $W$ could provide, and the e-h pair therefore behaves essentially as a free, unbound excitation \cite{onida2002electronic}. In this regime, the screened attraction has little effect on the spectrum, and the high-energy range is well described at the RPA level.

A key parameter in BSE+ is the size of $\mathcal T$. In the formal limit where $\mathcal{T}$ encompasses all e-h transitions, BSE+ reduces exactly to BSE, as illustrated for bulk systems in \cite{bseplus}. In practice, this limit is unreachable, and the central practical question is whether $\mathcal T$ can be made large enough to capture the energy range over which excitonic effects are significant, while remaining within computational reach. Like BSE, BSE+ cannot describe excitonic features arising from transitions outside $\mathcal{T}$, so if these extend beyond the chosen e-h subspace, the imaginary part of the spectrum will be incomplete in that range. A related issue arises in the transition region between the BSE-dominated low-energy part and the RPA-dominated high-energy part of the spectrum. In BSE+, the irreducible response switches abruptly from including $W$ to neglecting it at the boundary of $\mathcal{T}$, and spectral features in this intermediate range can carry artifacts from this truncation. 

In practice, however, these requirements are easily met for the materials considered in this work, and $\mathcal{T}$ can be chosen so that the transition region falls in a spectrally uninteresting range. BSE+ then delivers a spectrum that recovers the excitonic features at the BSE level at low energies while simultaneously capturing the plasmonic features at the RPA level at higher energies.

\subsection{Frequency-dependent broadening}
To better reproduce the experimental EELS spectra, we employ a frequency-dependent broadening parameter,

\begin{equation}
    \eta(\omega) = \eta_0 +  \eta_{2} \omega^2 
\label{eq:eta}
\end{equation}

where $\eta_0$ is a constant offset and $\eta_2$ governs the quadratic growth with frequency. In general, the broadening $\eta$ is proportional to the inverse quasi-particle lifetime, $\eta \propto 1/\tau$, which in many-body perturbation theory is set by the imaginary part of the self-energy $\eta\propto \Im \Sigma$. The frequency dependence and form of Eq. (\ref{eq:eta}) is motivated by Fermi liquid theory, in which the quasi-particle lifetime shortens as the excitation energy increases away from the Fermi level, due to the growing phase space available for electron-electron scattering. Specifically, the imaginary part of the GW self-energy scales as $\Im \Sigma \propto (\varepsilon - \varepsilon_F)^2$ \cite{bruus}. In a semiconductor, the analogous reference is the band edge, so that the broadening of a transition grows with its energy above the gap. Other scattering mechanisms contribute broadenings that are approximately energy-independent and are therefore absorbed into $\eta_0$. The form of Eq. (\ref{eq:eta}) then naturally captures the observed behavior, with sharp excitonic peaks at low frequencies where quasi-particle lifetimes are long, and progressively smoother features at larger frequencies. 

\section{Computational details}
\label{sec:computational_details}
We computed the density response function for a set of transition metal dichalcogenides (TMDCs) in the 2H phase: MoS$_2$, WSe$_2$, WS$_2$, MoTe$_2$ and MoSe$_2$. These TMDCs are widely studied 2D semiconductors with experimental EELS data available for benchmarking. The relaxed structures were obtained from the Computational 2D Materials Database (C2DB) \cite{c2db_1, c2db_2}. Ground-state DFT calculations were performed using an 800 eV plane-wave energy cutoff, along with the PBE exchange-correlation functional\cite{pbe}, and a $\Gamma$-centered Monkhorst-Pack $k$-point mesh corresponding to 70 $k$-points per $\mathrm{\AA}^{-1}$. This density is well above what is required for ground-state convergence and was chosen to provide a flexible starting point for the coarser meshes used in the subsequent response function calculations.

Two subsequent calculations with fixed electron density were performed to obtain the single-particle wave functions and eigenvalues required for the response function calculations. A coarser $k$-point mesh was used for the BSE $\tilde\chi^{\text{irr}}$ calculations, with material-specific densities listed in the supplementary material. These densities were chosen based on a convergence test performed on MoS$_2$, which indicated that a density of around 25-30 $k$-points per $\mathrm{Å}^{-1}$ is sufficient for the BSE calculation. The precise value for each material was selected as a compromise between covering the relevant energy range and staying within memory constraints. A finer $k$-point mesh was used for the converged non-interacting density response function $\chi^0$. 

For each material, one or two values of momentum transfers $q$ were selected based on the availability of experimental EELS data for benchmarking. For MoS$_2$, MoTe$_2$ and MoSe$_2$, the $q$-value used in the low-energy region differs from that used in the high-energy region. For MoS$_2$ and MoSe$_2$, low-energy data at the $q$-value of the high-energy panel were available, but did not resolve the excitonic features. We therefore compare to low-energy data at a smaller $q$ as indicated. For MoTe$_2$, we could only find low-energy experimental data that were not $q$-resolved, and which we compared to a $q=0\,\mathrm{Å}^{-1}$ calculation. Since $\qqq$ must lie on the BSE $k$-point grid, and the experimental $\qqq$ cannot be expected to fall exactly on the grid, the closest available $k$-point was used instead. To ensure that this same $\qqq$ was also represented on the $\chi^0$ grid, so that the BSE and RPA calculations were evaluated at identical momentum transfers, the $\chi^0$ $k$-point grid was chosen to be 2 or 3 times denser than the BSE grid in each in-plane direction. 

For all materials, a broadening of $\eta_0 = 40$ meV and $\eta_2 = 4 \cdot 10^{-3}$ eV$^{-1}$ was used. These values were empirically chosen on a single material (WSe$_2$) to retain the sharpness of the low-energy exciton peaks while smoothing out the rapid oscillations at higher frequencies, and were found to work equally well for the other materials without further adjustment. A plane-wave cutoff of 50 eV for the pair densities $\rho(\GG)$ was found sufficient to converge local field effects (reduced to 40 eV for MoTe$_2$ due to memory constraints). 

Bands within $\Delta E_B$ on each side of the band gap were included in the $\tilde{\chi}$ calculations, covering all transitions up to $E^{\text{QP}}_{\text{gap}}+\Delta E_B$. A default value of $\Delta E_B = 2$ eV was adopted and reduced for materials where the corresponding band count led to memory overflow. The values of $\Delta E_B$ for each material are available in the supplementary material. Spin-orbit coupling was included non-self-consistently in the BSE calculations, which is essential for an accurate description of the excitonic structure in 2D materials containing heavy elements. 

Since PBE is known to underestimate band gaps, and obtaining accurate quasi-particle band gaps is outside the scope of this work, a scissors shift was applied to the response functions by rigidly shifting the conduction bands. The shift was determined by aligning the lowest BSE exciton peak of the calculated spectrum to the corresponding experimental peak, and the same shift was applied consistently across RPA, BSE, and BSE+ calculations. The resulting quasi-particle direct band gap for each material is listed in the supplementary material.

The RPA, BSE, and BSE+ density response functions were then calculated using GPAW \cite{gpaw} with a truncated Coulomb kernel. The EELS spectrum was obtained from the density response function via Eq. (\ref{eq:eels_def}).

\section{Results}
\begin{figure}[!t]
\centering
\includegraphics[height=0.838\textheight]{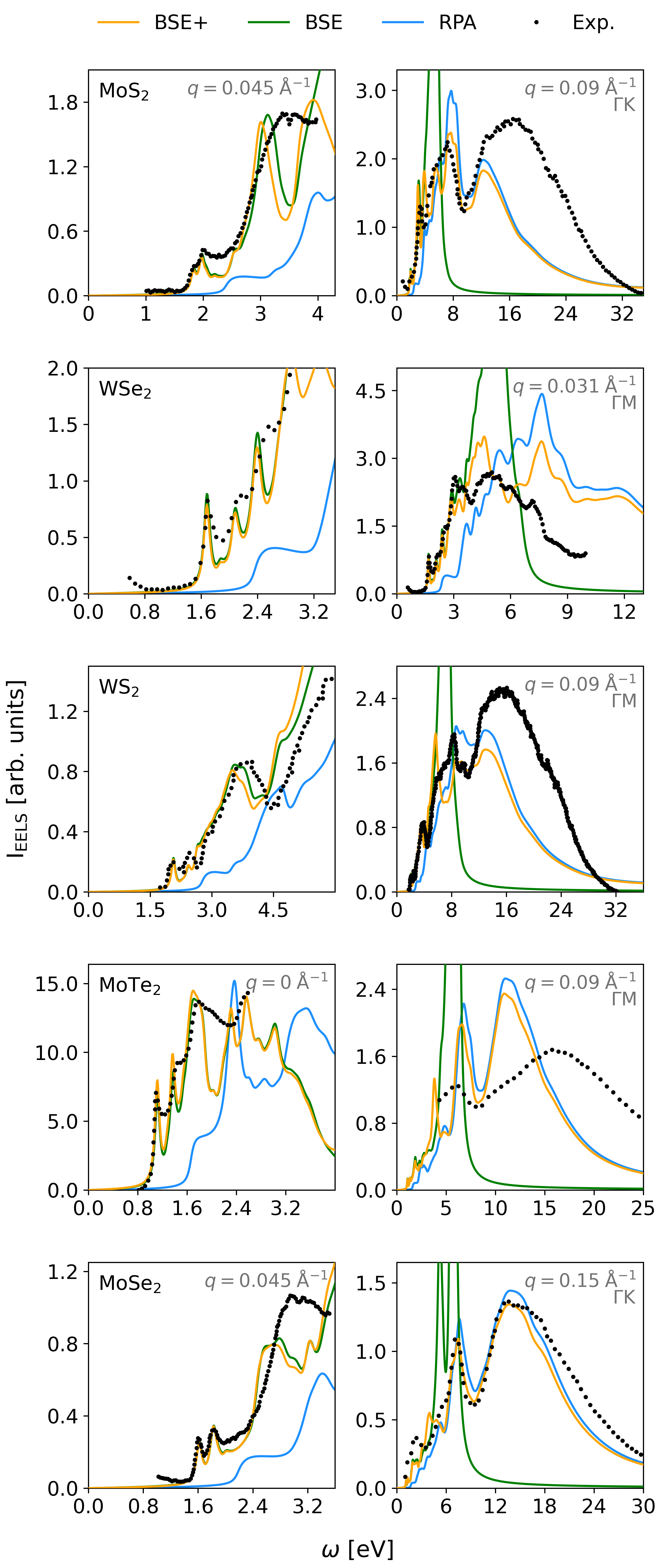}
\caption{EEL spectra calculated with BSE+ (orange), BSE (green), and RPA (blue). The experimental EELS data are marked with black dots, and are obtained from \cite{MoS2_MoSe2_WS2_small_w, MoTe2_MoS2_exp_data_large_w} ($\text{MoS}_2$), \cite{WSe2_exp_data} ($\text{WSe}_2$), \cite{MoS2_MoSe2_WS2_small_w} ($\text{WS}_2$), \cite{MoTe2_exp_data_small_w, MoTe2_MoS2_exp_data_large_w} ($\text{MoTe}_2$), and \cite{MoS2_MoSe2_WS2_small_w} ($\text{MoSe}_2$). The targeted $q$-values are indicated in each panel. Details of their selection are given in Sec. \ref{sec:computational_details}. Experimental data were scaled by a material-specific constant to match the height of the first calculated BSE peaks, which is valid since the unit of the y-axis is arbitrary.}
\label{fig:eels}
\end{figure}

To assess the performance of BSE+ for 2D materials, we calculate the EEL spectra with BSE+, BSE, and RPA and compare them to experimental data in Fig. \ref{fig:eels}. In all cases, BSE+ captures the low-energy excitonic features at the level of the BSE, while reproducing the high-energy plasmonic structure at the level of the RPA. Thus, BSE+ yields good agreement with experiment across the full energy range and at a computational cost comparable to that of the BSE. 

As also observed for bulk systems \cite{bseplus}, BSE+ does not generally improve the position or shape of individual peaks beyond that provided by the underlying BSE and RPA.  When either method misplaces a feature, BSE+ inherits the same error. In WSe$_2$, for example, the experimental intensity decreases above 5 eV, while the RPA and BSE+ have not yet begun to decay to zero in this range. Similarly, in MoTe$_2$, the experimental plasmon near 15 eV is predicted by RPA to be around 11 eV, and the same behavior is predicted by BSE+. In MoS$_2$ and MoSe$_2$, the two lowest-energy exciton peaks are well described by the BSE, but the position of the next peak is underestimated, and again BSE+ does not correct this.

Beyond these inherited limitations, a more striking shortcoming of the BSE itself becomes visible at finite $q$, where it produces a spurious plasmon peak just before the spectrum decays to zero. Although EELS probes only the imaginary part of $\chi$, this quantity depends on both the real and imaginary parts of $\Tilde{\chi}^{\text{irr}}$ through the Dyson equation. Schematically, a plasmon arises at the point at which the Dyson denominator vanishes, i.e., where $1 - V_{\text{trunc}}\,\text{Re}(\Tilde{\chi}^{\text{irr}})$ has an eigenvalue of zero. In BSE, the irreducible response function is constructed from a truncated transition space, and the missing high-energy transitions cause its real part to be underestimated. The Dyson denominator, therefore, vanishes at an unphysical energy that depends on the size of the truncated transition space, producing a spurious collective mode in the BSE spectrum.

The spurious plasmon is a generic shortcoming of the BSE, and it is present in both bulk and 2D systems whenever the transition space is not large enough to converge the real part of the density response function across the full energy range. As shown in \cite{bseplus}, this is also a significant issue for bulk semiconductors and insulators, and the same mechanism is at play in the 2D materials studied here. BSE+ resolves this by restoring the missing transitions at the RPA level in the irreducible density response function, so that the Dyson denominator vanishes only when it physically should, and the spurious plasmon is strongly suppressed or even absent in the BSE+ spectra.

\begin{figure}[!h]
\centering
\includegraphics[width=\linewidth]{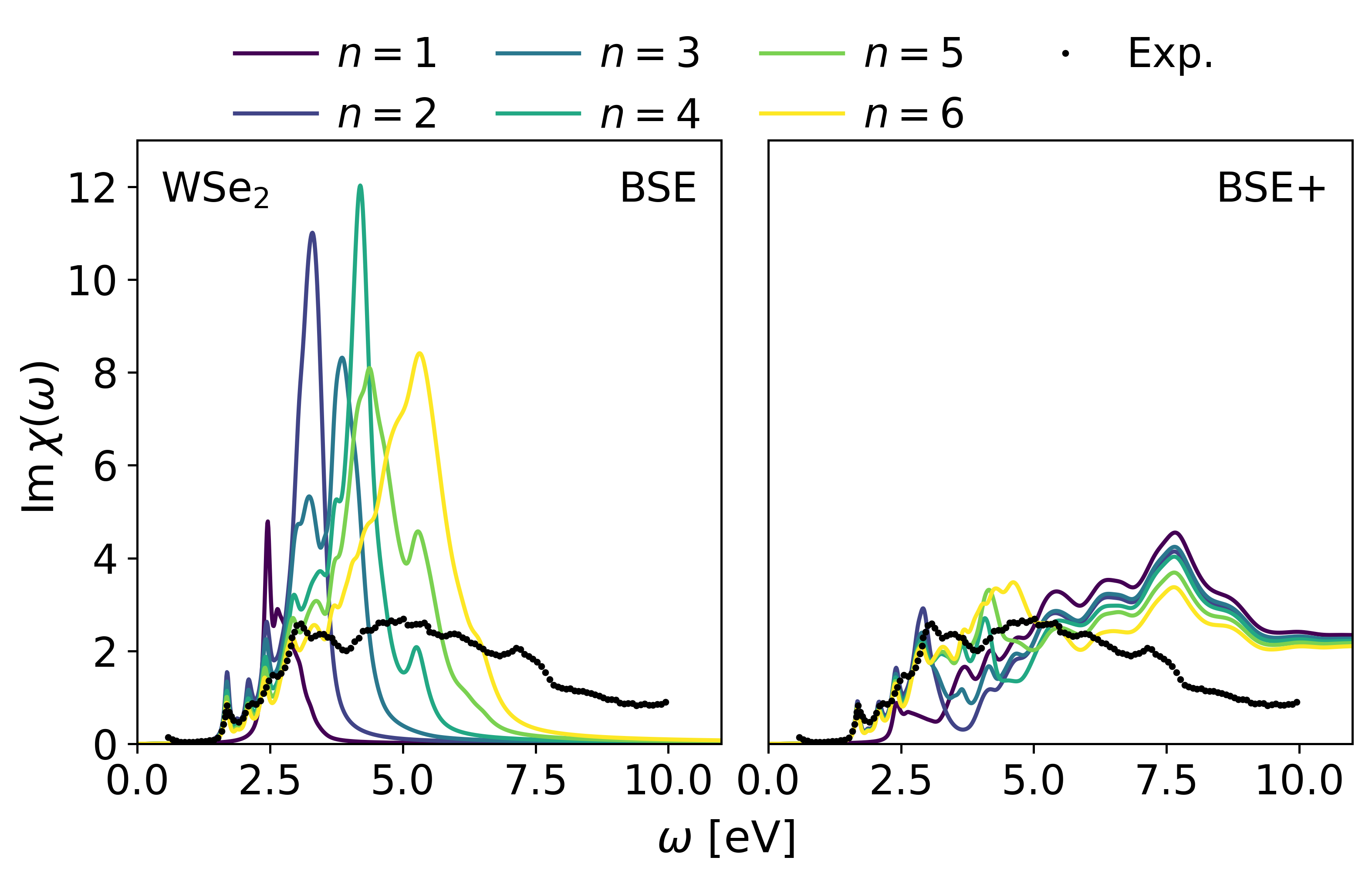}
\caption{Convergence of BSE and BSE+ with respect to the number of bands in the BSE calculation. The number $n$ indicates that there are $n$ valence and $n$ conduction bands included in $\mathcal{T}$. Note that for odd $n$, only one member of a spin-orbit coupled band pair is included, which would not be appropriate for a production calculation. Such configurations are included here to more densely sample the convergence behavior.}
\label{fig:conv}
\end{figure} 

Fig. \ref{fig:conv} illustrates this convergence behavior explicitly for WSe$_2$, where we gradually increase the number of valence and conduction bands included in the transition space $\mathcal{T}$. In BSE, the spurious plasmon shifts toward higher energies as more bands are added, but persists throughout the range of $\mathcal{T}$ accessible within our computational resources. In BSE+, only a small remnant of the spurious plasmon in the BSE is visible, and its dependence on the size of $\mathcal{T}$ is much weaker. With as few as two valence and two conduction bands, BSE+ already provides a good description of the full spectrum, which further improves as more bands are added. This mirrors the convergence behavior reported for bulk systems in \cite{bseplus}, and confirms that neglecting only the e-h attraction in transitions outside $\mathcal{T}$, rather than the transitions themselves, translates directly into faster and more controlled convergence. 

This remnant of the BSE spurious plasmon is not unique to WSe$_2$. It is also visible in, e.g., the WS$_2$ spectrum in Fig. \ref{fig:eels}, where BSE+ exhibits a small peak near 5 eV with no counterpart in the experimental data. The feature originates in the same mechanism that produces the spurious BSE plasmon, and is suppressed but not entirely eliminated by the RPA-level correction. Enlarging $\mathcal{T}$ would push the remnant to higher energies and reduce its amplitude, which is consistent with the convergence behavior in Fig. \ref{fig:conv}. Despite this residual artifact, the overall agreement of BSE+ with the experimental EEL spectra across the full energy range is markedly better than that achieved by BSE or RPA individually.

\section{Conclusion}
In conclusion, we have presented a detailed account of the BSE+ method and extended it to 2D materials. By including transitions outside the active BSE e-h subspace at the RPA level in the irreducible density response function, BSE+ achieves a much better convergence of the real part of the density response function, without increasing the computational cost relative to standard BSE. Applied to a set of TMDC monolayers, BSE+ reproduces the excitonic features at the BSE level at low energies, while simultaneously capturing the plasmonic structure at the RPA level at higher energies, which is in good agreement with experimental EEL spectra across the full energy range. The spurious plasmon present in BSE spectra is strongly suppressed or absent in the BSE+ spectra. By yielding converged density response functions across the entire spectrum, BSE+ paves the way for applications that require the full energy range, such as total-energy calculations via the adiabatic connection fluctuation-dissipation theorem.

\section{Acknowledgements}
The authors thank Prof. Kazu Suenaga and Dr. Jinhua Hong (SANKEN, The University of Osaka) for sharing their experimental EELS data. The authors acknowledge funding from the Villum Investigator Grant No. 37789 supported by VILLUM FONDEN and from the Novo Nordisk Foundation Data Science Research Infrastructure 2022 Grant:  A high-performance computing infrastructure for data-driven research on sustainable energy materials, Grant no. NNF22OC0078009.

\bibliography{bibliography} 
\end{document}